\documentstyle[12pt]{article}
\begin{document}
\pagestyle{empty}
\begin{flushright}
{BROWN-HET-907} \\
{May 1993}
\end{flushright}
\vspace*{5mm}
\begin{center}
{\bf A Nonsingular Two Dimensional Black Hole} \\
[10mm]
M. Trodden$^{(1)}$, V.F. Mukhanov$^{(2)*}$\footnotetext{On leave of absence
from Institute for Nuclear Research, Academy of Sciences, 117312 Moscow,
Russia.}, R.H. Brandenberger$^{(1)}$ \\
[10mm]
(1) Department of Physics \\
    Brown University \\
    Providence RI. 02912. \\
    USA. \\[5mm]

(2) Institut f\"ur Theoretische Physik \\
    ETH Z\"urich, Honggerberg \\
    CH-8093 Z\"urich, Switzerland. \\[2cm]

{\bf Abstract}
\end{center}
\vspace*{3mm}
We construct a model of gravity in 1+1 spacetime dimensions in which the
solutions approach the Schwarzschild metric at large $r$ and de Sitter space
far inside the horizon. Our model may be viewed as a two dimensional
application of the `Limiting Curvature Construction' of reference\cite{M&B 92}.
\\

\newpage\setcounter{page}{1}\pagestyle{plain}

\section{Introduction}
It has long been realised that the well known singularity theorems of Penrose
and Hawking\cite{P&H} imply that general relativity is an incomplete
description of the behaviour of spacetime at high curvatures. It is commonly
believed that the successful quantization of gravity will provide us with the
modifications to the theory that are neccessary to avoid the prediction of a
geodesically incomplete spacetime manifold. However, as yet such a quantum
theory of gravity cannot be said to exist and even if and when it does we will
undoubtably need to look to effective theories in order to understand its
implications\cite{B&V 89}. Since a generic prediction of all attempts to
construct a consistent theory of quantum gravity is that the Einstein-Hilbert
action be modified at high curvatures by extra, usually higher derivative,
invariant terms\cite{S&S 74}-\cite{H&V 74} we may expect that this structure
will be represented in any effective theory.

We may therefore ask whether it is possible to construct an effective theory of
gravity that is intermediate between General Relativity and Quantum Gravity,
ie. one that (a) gives the correct (Einstein) behaviour in the low curvature
limit and (b) encompasses the nonsingular nature that we hope to be contained
in our full theory.

A class of such theories have been studied in references\cite{M&B 92}-\cite{B&T
93} with regard  to a variety of spacetimes in 3+1 dimensions. The resulting
field equations obey the requirements above and in particular satisfy (b) by
forcing the spacetime to become de Sitter (and thus singularity free) at high
curvatures. This method is known as the `Limiting Curvature
Construction'\cite{GMF 88}. However, whilst the theories are successful the
calculations can be formidable and numerical solutions are required since the
structure of the equations is highly complex.

As a toy model, therefore, it may be instructive to consider the same type of
theory in 1+1 dimensions. Our motivations are threefold: (i) to make clear the
structure of the 3+1 dimensional models by analogy, (ii) to investigate the
relationship between our work and string theory motivated black hole
studies\cite{G&M 88}-\cite{CGHS 92} and (iii) to explore new ideas in a simple
and more hospitable setting. Of course, success in two dimensions does not
imply success in four but we may hope to shed some new light on the more
complicated models with this method.

In this paper we present a modified action for 1+1 dimensional gravity which we
use to describe a black hole spacetime. We show that away from regions of high
curvature we recover the Schwarzschild spacetime and that as we approach the
high curvature regions the spacetime becomes de Sitter. All our solutions are
analytic and the structure of the equations is easy to see at all times. As in
the four dimensional models we achieve the nonsingular behaviour by
parametrizing the higher derivative action by means of a nondynamical scalar
field and choosing its potential energy such that the curvature remains
bounded.

As we mentioned above, one motivation for this study is the current interest in
two dimensional black holes in string theory\cite{W 91}-\cite{H&S 92}. There
has recently been substantial progress towards nonperturbative solutions of two
dimensional string theory. Of particular interest have been solutions
corresponding to black holes\cite{H 92}. One approach to solving this
problem\cite{CGHS 92} is to consider an effective action for the massless modes
of the string, in particular the graviton and the dilaton (the tachyon is
usually set to zero), and to study black hole solutions in this theory.

A further motivation for studying dilatonic gravity in two dimensions is the
hope that such a study might reveal the answer to the quantum mechanical
information loss problem of black holes. What happens to the information which
has entered the Schwarzschild radius before the black hole evaporates
completely by Hawking radiation\cite{H&S 92}? Does quantum mechanics break down
in the sense that a pure state evolves into a mixed state, is information
contained in the final stages of Hawking radiation, or do stable black hole
remnants remain? In four dimensions, the calculational techniques break down
during the final stages of Hawking radiation, and we are unable to answer the
above questions. However, in a simple two dimensional toy model we might be
able to determine the answer using analytical means.

\section{The Field Equations}
We will consider a Lagrangian consisting of a potential for our non-dynamical
scalar field and a general coupling to the Ricci scalar: (for convenience we
adopt the notation of Banks and O'Loughlin\cite{B&O 91})

\begin{equation}
{\cal L} = \sqrt{-g}\, (V(\phi) + D(\phi)R)
\end{equation}
In two spacetime dimensions, the most general renormalizable Lagrangian for a
graviton and dilaton field theory can be put into this form via a conformal
transformation.

Variation of the action with respect to the scalar field yields

\begin{equation}
- \sqrt{-g}\, \frac{\partial V}{\partial \phi}(\phi) =
 \frac{\partial D}{\partial \phi}(\phi) \sqrt{-g}\, R
\end{equation}
Now consider variations with respect to the metric
\begin{eqnarray}
\delta S & = & \int d^{2}x \ [(V(\phi) + D(\phi)R)\, \frac{1}{2}\,
               g_{\alpha\beta} \sqrt{-g}\, \delta g^{\alpha\beta}  +
               \nonumber \\
         &   &  \sqrt{-g}\, (-D(\phi)R_{\alpha\beta}\delta g^{\alpha\beta}-
               D(\phi)\nabla^{2} g_{\alpha\beta} \delta g^{\alpha\beta} +
               D(\phi) \nabla_{\alpha}\nabla_{\beta} \delta
               g^{\alpha\beta}) ]
\end{eqnarray}
Integrating by parts twice, ignoring surface terms and requiring $\delta S = 0$
gives

\begin{equation}
-D(\phi)R_{\alpha\beta} + (\nabla_{\alpha}\nabla_{\beta} - g_{\alpha\beta}
\nabla^{2})D(\phi) + \frac{1}{2} V(\phi)g_{\alpha\beta} +
\frac{1}{2} D(\phi)Rg_{\alpha\beta} = 0
\end{equation}
Finally we note that in two dimensions $R_{\alpha\beta}=g_{\alpha\beta}R/2$
which puts our field equations in the form

\begin{equation}
V(\phi)g_{\alpha\beta} = 2(\nabla^{2}g_{\alpha\beta} -
\nabla_{\alpha}\nabla_{\beta})D(\phi)
\end{equation}
Equations (2) and (5) are our vacuum field equations. Together they form three
equations, only two of which are independent.

\section{Two Dimensional Black Hole}
By redefining $\phi$ we can set $D(\phi)=1/\phi$. In two spacetime dimensions
we can for any `static' metric chose a gauge in which
\begin{equation}
g_{\mu\nu} = {\rm diag}(-f(r)\ ,\ f(r)^{-1})
\end{equation}
Our field equations now become

\begin{equation}
\frac{\partial V}{\partial \phi}(\phi) = \frac{1}{\phi^{2}}R
\end{equation}
\begin{equation}
g_{\alpha\beta}V(\phi) = 2(g_{\alpha\beta}\nabla^{2} -
                           \nabla_{\alpha}\nabla_{\beta})\frac{1}{\phi}
\end{equation}
Expanding the covariant derivative in the second expression we obtain

\begin{equation}
g_{\alpha\beta} V(\phi) = 2g_{\alpha\beta}g^{\gamma\lambda}(\partial_{\gamma}
              \partial_{\lambda}
             -\Gamma^{\sigma}_{\gamma\lambda}\partial_{\sigma})\frac{1}{\phi} -
             2(\partial_{\alpha}\partial_{\beta} -
              \Gamma^{\gamma}_{\alpha\beta}\partial_{\gamma})\frac{1}{\phi}
\end{equation}
First we consider the time-time component. This gives us

\begin{equation}
\phi^{3}V(\phi) + 2f\phi\phi'' - 4f{\phi'}^{2} +
                        f'\phi\phi' = 0
\end{equation}
where we have written $()' \equiv \partial ()/\partial r$. Similarly the
space-space equation becomes

\begin{equation}
\phi^{2} V(\phi) +f'\phi' = 0
\end{equation}
and the variational equation for the scalar field is

\begin{equation}
\frac{\partial V}{\partial \phi}(\phi) = \frac{1}{\phi^{2}}(-f'')
\end{equation}
Writing $\Phi \equiv \phi'/\phi$ and combining equations (11) and (12) we
obtain the following differential equation

\begin{equation}
\frac{\Phi'}{\Phi} - \frac{\phi'}{\phi} = 0
\end{equation}
which yields, finally

\begin{equation}
\frac{\phi'}{\phi^{2}} = - A,
\end{equation}
where $A$ is a constant of integration. So we may write, formally

\begin{equation}
\phi = \frac{1}{Ar+B}
\end{equation}
Note that by shifting the $r$ coordinate we can without loss of generality
assume $B=0$.

Now that we have obtained the dynamics of the scalar field as a function of the
metric our logic will be as follows:

\begin{enumerate}
\item Choose $f(r)$ to be of the Schwarzschild form in the asymptotic region $r
\rightarrow \infty$ ($\phi \rightarrow 0$) and solve exactly for the field
$\phi$ and the potential $V(\phi)$.

\item Choose the potential $V(\phi)$ such that at large $\phi$ (small $r$) the
curvature $R$ is bounded.

\item Construct an interpolating potential $V(\phi)$ with the correct behaviour
in both asymptotic regions.

\item Use the interpolating potential to solve exactly for the general form of
$f(r)$.
\end{enumerate}

Note that in two dimensions, $R$ is the only curvature invariant. Hence, after
bounding $R$, all invariants will be bounded. We will show that this procedure
singles out de Sitter space as the unique solution. This is a substantial
simplification compared to the four dimensional case, where it is necessary to
introduce higher derivative curvature invariants in the Lagrangian in order to
obtain de Sitter space as a solution.

In the asymptotic region $r \rightarrow \infty$ we have

\begin{equation}
f(r) = 1 - \frac{2m}{r}
\end{equation}
We then obtain, from equation (12) with $\phi \rightarrow 0$

\begin{equation}
V(\phi) \sim 2m A^{3} \phi^{2}(r)
\end{equation}
where we have ignored a constant of integration.

For $\phi \rightarrow \infty$ ($r \rightarrow 0$) we demand that $R$ remains
bounded. From (7) it follows that
$\frac{\partial V}{\partial \phi} \sim \phi^{-2}$ and hence, introducing a new
proportionality constant $l$,

\begin{equation}
V(\phi) = \frac{2}{l^{2}}\frac{1}{\phi}
\end{equation}
By integrating (11) we obtain

\begin{equation}
f(r) = \frac{r^{2}}{l^{2}} - C
\end{equation}
and by rescaling $r$ we can take $C=1$. In Appendix A it is shown that the
metric given by (6) and (19) corresponds to de Sitter space.

As an example, we consider a specific potential which interpolates between (17)
and (18), given by

\begin{equation}
V(\phi) = \frac{2mA^{3}\phi^{2}}{1 + mA^{3}l^{2}\phi^{3}}
\end{equation}
Using the above form of the potential in equation (11) we arrive at an exact
form for the metric component $f(r)$.
Using $\phi(r)= 1/Ar$, equation (11) becomes

\begin{equation}
f'(r) = \frac{2mr}{r^{3} + ml^{2}}
\end{equation}
which can be integrated exactly between $r_{0}$ and $r$ to give

\begin{eqnarray}
f(r) & = & \frac{1}{3} \left(\frac{m}{l}\right)^{2/3}
\ln\left[\frac{r^{2}-(ml^{2})^{1/3}r+(ml^{2})^{2/3}}
              {r_{0}^{2}-(ml^{2})^{1/3}r_{0}+(ml^{2})^{2/3}}
   \left(\frac{r_{0}+(ml^{2})^{1/3}}{r+(ml^{2})^{1/3}}\right)^{2}\right] +
\nonumber \\
     &   & \frac{2}{\sqrt3}\left(\frac{m}{l}\right)^{2/3}
\left[\arctan\left(\frac{2r-(ml^{2})^{1/3}}{\sqrt{3}\,(ml^{2})^{1/3}}\right)
\right.- \nonumber \\
     &   &
\left.\mbox{\hskip2.4in}\arctan\left(\frac{2r_{0}-(ml^{2})^{1/3}}{\sqrt{3}\,(ml^{2})^{1/3}}
\right)\right]
\end{eqnarray}
This function is plotted in figure 1.

We have now completed our calculation. The two dimensional spacetime that we
have constructed is defined by the metric (6) with $f(r)$ given by the above
expression. At macroscopic distances from $r=0$ the spacetime is
indistinguishable from the Schwarzschild spacetime which would be predicted by
general relativity. As we approach the region of high curvature non-Einstein
effects become important and the transition is smoothly made to a de Sitter
spacetime. This model is singularity free while possessing all the relevant
features of the usual Schwarzschild black hole.

\section{Conclusions}

We have constructed a class of $1 + 1$ dimensional gravity theories which have
nonsingular black hole solutions. The models are obtained using the `Limiting
Curvature Construction' of reference \cite{M&B 92} . Starting from the general
higher derivative gravity action parametrized by means of a nondynamical scalar
field $\phi$, our subclass of models is obtained by choosing the potential
$V(\phi)$ such that the Ricci curvature is bounded. This leads to the
asymptotic condition (18) for $V(\phi)$ in the large $\phi$ limit.

Any potential satisfying (18) for large $\phi$ leads to a theory in which
space-time approaches de Sitter space at large curvatures. Choosing an
appropriate form of $V(\phi)$ for small $\phi$ (see (17)) leads to a model in
which the `static' solutions approach the `usual' Schwarschild metric at large
distances.

Note that the singularity-free nature of our solutions emerges as a consequence
of the `Limiting Curvature Principle' and is not put in by hand.

The principles of our construction differ from those upon which the recent
construction of nonsingular black holes by Banks and O'Loughlin \cite{B&O 92}
are based. These authors also obtain solutions which have a horizon and
approach de Sitter space inside the horizon, but their solutions are different
outside.
Our black holes have a dilaton field outside of the Schwarschild horizon which
rapidly approaches zero, and thus they for large $r$ correspond to the
dimensionally reduced four dimensional black hole. Note that the black hole
solutions of Ref. \cite{W 91} have a nonvanishing dilaton at large $r$.

Our theory should be seen as a two dimensional effective theory of gravity
which contains the appropriate Schwarzschild solution as a unique low curvature
approximation while at the same time encapsulating the singularity free nature
which we hope will be a feature of a more complete, quantum, theory of gravity.

\centerline{\bf Acknowledgements}

We thank Andrew Sornborger for helpful discussions. This work has been
supported in part by the US Department of Energy under Grant DE FG02-91ER40688,
Task A. V.M. is supported by the Schwiss National Foundation.

\appendix
\section{The 2-D De Sitter Spacetime}
Here we will show that the two dimensional de Sitter spacetime may be
represented in the form (6). Consider the metric (6) with

\begin{equation}
f(r) = \left(\frac{r}{l}\right)^{2} - 1
\end{equation}
Let us introduce a new coordinate $\tau$ defined by

\begin{equation}
\frac{d\tau}{dr} = \left[\left(\frac{r}{l}\right)^{2}-1\right]^{-1/2}
\end{equation}
Solving for $\tau(r)$ we obtain

\begin{equation}
\frac{\tau(r)}{l} = {\rm arccosh}\left(\frac{r}{l}\right)
\end{equation}
which implies that our line element in $(t,\tau)$ coordinates is

\begin{equation}
ds^{2} = d\tau^{2} - \sinh^{2}(\tau/l) \,dt^{2}
\end{equation}
which is of the form of a de Sitter spacetime with a de Sitter bounce as
required.

\bigskip
\centerline{\bf Figure Caption}

\noindent{\bf Figure 1:} The metric coefficient $f(r)$ as a function of $r$.
\end{document}